\documentclass[remotesensing,article,accept,moreauthors,pdftex]{Definitions/mdpi} 

\firstpage{1} 
\pubvolume{1}
\issuenum{1}
\articlenumber{0}
\pubyear{2021}
\copyrightyear{2020}
\datereceived{} 
\dateaccepted{} 
\datepublished{} 
\hreflink{https://doi.org/} 

\Title{Snowcover Survey over an Arctic Glacier Forefield: Contribution of Photogrammetry 
to Identify ``Icing'' Variability and Processes}

\Author{Eric Bernard$^{1,\ddagger,*}$\orcidA{}, Jean-Michel Friedt$^{2,\ddagger,*}$\orcidB{} and Madeleine Griselin$^{1,\ddagger}$}

\AuthorNames{Eric Bernard, Jean-Michel Friedt, Madeleine Griselin}

\address{%
$^{1}$ \quad CNRS / Th\'eMA, University of Bourgogne Franche Comt\'e, France ; eric.bernard@univ-fcomte.fr\\
$^{2}$ \quad FEMTO-ST, CNRS/University of Bourgogne Franche Comt\'e, France ; jeanmichel.friedt@femto-st.fr}

\corres{Correspondence: jmfriedt@femto-st.fr, eric.bernard@univ-fcomte.fr }

\secondnote{These authors contributed equally to this work.}

\abstract{Current climate shift has significant impacts on both quality and quantity of snow precipitation. 
This directly influences spatial variability of the snowpack as well as cumulative snow height. Contemporary 
glacier retreat reorganizes periglacial morphology: while the glacier area decreases, the moraine area increases. 
The latter is becoming a new water storage potential almost as important as the glacier itself, but with a much 
more complex topography. {\color{red}This work hence fills one of the missing variables of the hydrological budget equation
of an arctic glacier basin by providing an estimate of the snow water equivalent (SWE) of the moraine contribution. Such a 
result is achieved by} 
investigating {\color{red}Structure from Motion (SfM) image processing applied to pictures 
collected from an Unmanned Aerial Vehicle (UAV)} as a method to produce snow depth maps over the proglacial moraine area. 
Several UAV campaigns were carried out on a small glacial basin in Spitsbergen (Arctic): measurements were made 
at the maximum snow accumulation season (late April) while reference topography maps were acquired at the end of 
hydrological year (late September) when the moraine is mostly free of snow. 
Snow depth is determined from {\color{red}Digital Surface Model (DSM)} subtraction. Using dedicated and natural ground 
control points for relative positioning of the DSMs, the relative DSM georeferencing with sub-meter accuracy removes 
the main source of uncertainty when assessing snow depth. For areas where snow is deposited on bare rock surfaces, 
{\color{red}the} correlation between avalanche probe in-situ snow depth measurements and DSM differences {\color{red}is} excellent. 
Differences in ice covered areas between the two measurement techniques are attributed to the different quantities 
measured: while the former only measures snow accumulation, the latter includes all ice accumulation during winter 
through which the probe cannot penetrate, in addition to the snow cover. When such inconsistencies are observed, 
icing thicknesses are the source of the discrepancy observed between avalanche probe snow cover depth measurements and 
differences of DSMs.}

\keyword{Snowcover ; {\color{red}Snow water equivalent} ; Cryosphere ; Moraine ; Arctic ; UAV-SfM ; Spatial dynamics ; Photogrammetry}  

\begin{document}


\section{Introduction}
Cryosphere dynamics are highly dependent on snowcover processes, which trigger further hydrological processes. Snowmelt runoff is part 
of fresh water fluxes reaching oceans and is thus strongly linked with snowpack spatio-temporal variability over a season 
(\cite{Nuth2010}, \cite{Hock2005}). Furthermore, in environments such as mountainous regions, snowpack dynamics often dominate water storage and release (\cite{Lehning2006}) which strongly influences geomorphological adaptation.  
In high Arctic, year after year, a glacier retreat trend is generally observed while the {\color{red}area} of the proglacial 
moraine increases {\color{red}at} the same time (\cite{Bernard2016}, \cite{Tonkin2016}). Consequently, the corresponding snowpack surface on ice-free ground also becomes wider. With respect to a glacial hydro-system, {\color{black} this pro-glacial moraine area} should now be considered as a 
{\color{black} increasingly important} contributor to outflows {\color{black} in addition to} the glacier snowpack itself (\cite{Etzelmuller2010}, \cite{Martin-Espanol2015}). However, due to the glacier forefield topographical characteristics, snowpack in the moraine is much more challenging to monitor. Indeed, the micro and local rough topography result in a high degree of seasonal and inter-annual variability in spatial distribution (\cite{Derksen2000}). Snow banks and massive accumulations contrast with convex area, particularly eroded by the wind or influenced by black-body effect (\cite{Barr2014}).   

In addition to {\color{black} such considerations}, ongoing dynamics induced by climate shift imply an increase of short events {\color{red}with
long lasting consequences} such as rain on snow (\cite{Eckerstorfer2012}), wind effects (\cite{Svendsen2002}) or even sudden heavy snowfalls (\cite{Bednorz2011}). {\color{black} The occurrence of}
these phenomena is observed to increase {\color{black}over time}, strongly contributing to the modification of snowcover dynamics (\cite{Sobota2016}). 

In the specific case of a morainic structure, collecting snow {\color{black}cover} data that are representative of the spatial  distribution of snow
{\color{black}depth is challenging due to the topographic discontinuities}. Thus, remote sensing methods could be considered as an alternative or, better still, a complement to ground  observations. In recent years, the use of unmanned aerial vehicle (UAV) data acquisition has emerged as a well suited method for investigating geomorphological changes due to climate shift (\cite{Stumpf2015}, \cite{Cook2017}). Similarly, cryospheric processes can also be measured quite accurately (\cite{Bhardwaj2016}, \cite{Fonstad2013}, \cite{Lucieer2013}, \cite{Westoby2015}). According to these works, the use of combined UAV with Structure from Motion (SfM) data processing is well suited for glacial/periglacial environment (\cite{Nolan2015}, \cite{Gindraux2017}, \cite{Ryan2015}), and especially when 
follow{\color{black}ing} fast and short processes (\cite{Smith2015}, \cite{Eltner2016}) {\color{red}lasting a few hours to a few days, such as 
flash floods inducing large transfers of sediments and carving canyons in the moraine or washing in only a few hours the snow cover that took 
weeks to months to accumulate}. In addition, we showed in past works (\cite{Bernard2017}) that in Arctic, climatic conditions as well as harsh field campaigns need a flexible {\color{black} means for} carrying out a monitoring task{\color{red}: a UAV is deployed on short notice in less than a couple of hours, the 
time needed to reach any launch site in the glacier basin in addition to being granted flight permission}.
{\color{black}Due to quickly varying weather conditions, field campaigns should be carried out at short notice to meet the assumption of
Structure from Motion (SfM) processing of constant illumination and static terrain features (\cite{Jagt2015}).}
{\color{black} The weather changes {\color{red}on a daily basis with low cloud ceiling and strong winds preventing flight as well
as satellite imagery which will not penetrate clouds: planning needs to be adjusted on a daily basis as short as possible after the 
brief (heavy snow fall or rain, rain on snow) event occurred.} Thus field campaigns have to be carried out as fast as possible to ensure 
both data acquisition and data homogeneity. There is all the more reason to use and apply this workflow for snowpack survey even 
if it was shown that photogrammetry on snow remains challenging (\cite{Jagt2015}).}

Here, we investigate snowpack accumulation from one year to another time frame, within a small proglacial moraine. {\color{red}We aim at highlighting} the capability of UAV {\color{red}collected images
processed using SfM} to assess specific snowcover {\color{red}evolution} processes in a typical 
Arctic environment. In this work, this topic is mainly discussed through icing field dynamics. The occurrence of such phenomena has been described for several areas of Svalbard (\cite{Bennett2012}, \cite{Sobota2016}). According to \cite{BukowskaJania2005}, water storage and release during the winter reflects the development of the subglacial drainage system and its capacity in the cold season. Considering that icing fields are well described in the literature (\cite{Bukowska-Jania2007}, \cite{Evans2009}, \cite{Rutter2011a}, \cite{Blum2011}), our approach {\color{black}focuses on} seasonal evolutions and {\color{black}quantifying} water release due to icings disappearance.

\begin{figure}[h!tb]
\centering
	\includegraphics[width=\linewidth]{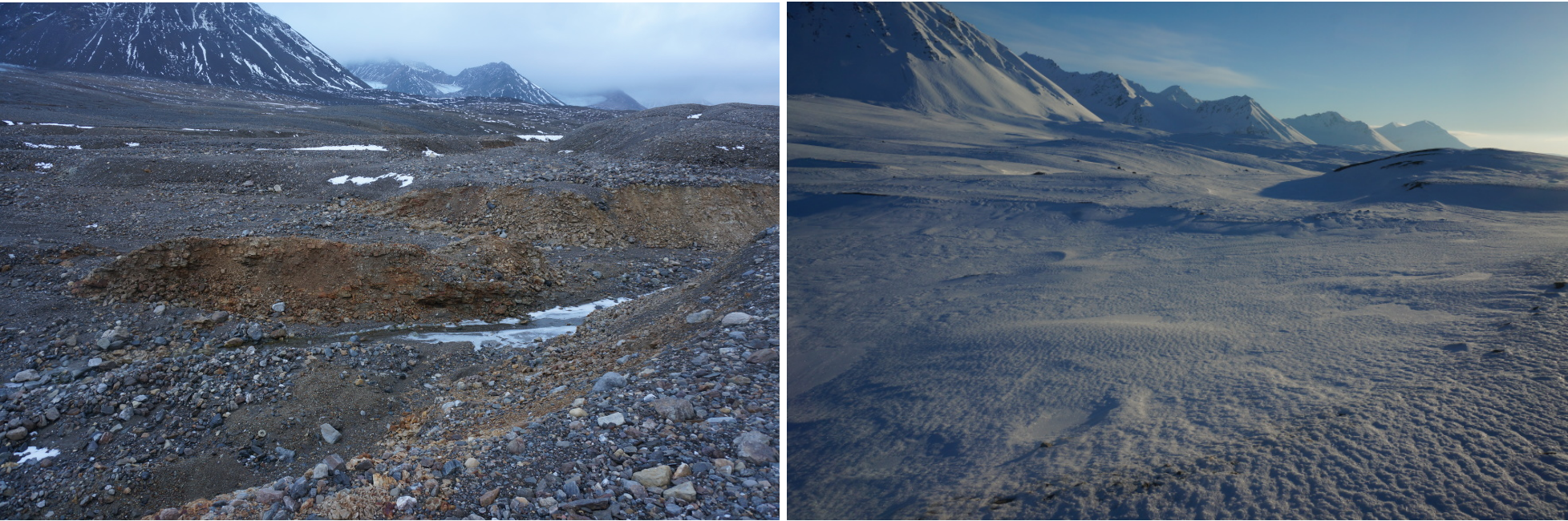}
\caption{Two pictures, taken in October 2016 (left) and April 2017 (right), from the same location, 
illustrating the snowpack distribution over the moraine topography.}\label{1}
\end{figure}

In this paper, the main purposes are as follow:
\begin{itemize}[leftmargin=*,labelsep=5.8mm]
\item	to derive {\color{red}Digital Surface Models (DSM)} at maximum/minimum snow accumulation in order to quantify the snow water equivalent (SWE);
\item	to analyze icing dynamics over Austre Lov\'en proglacial moraine by focusing on highly responsive areas such as river channels (Fig. \ref{1}).
\end{itemize}

\section{Study site and morphological characteristics}
\begin{figure}[h!tb]
\centering
	\includegraphics[width=\linewidth]{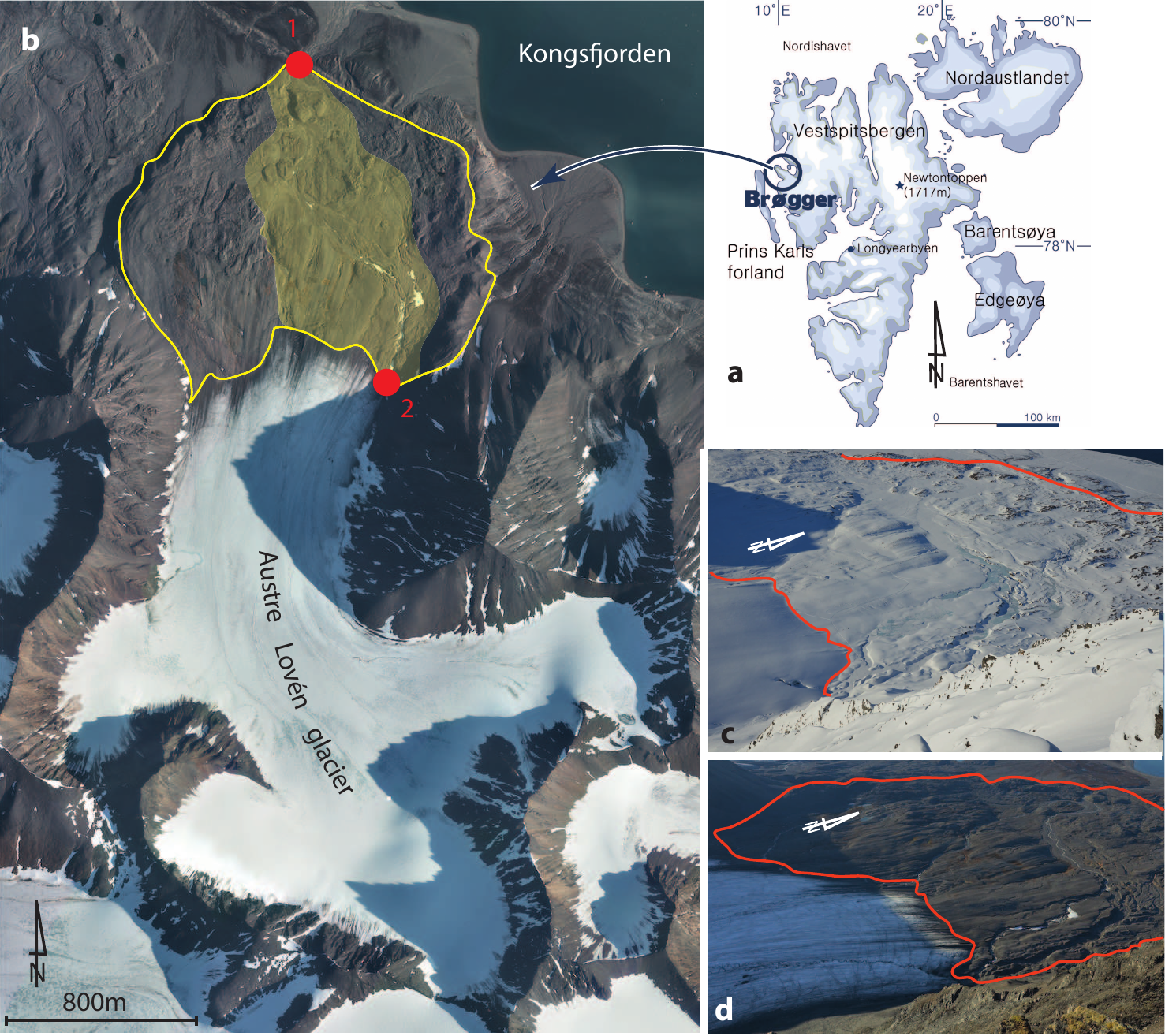}
\caption{General settings of the investigated area \textbf{a} with a focus (\textbf{b}) on Austre Lov\'en glacier basin. The proglacial moraine is delimited by the yellow line. The area where most of the hydrological and geomorphological processes occurred is in pale yellow, including the main water system, from the glacier outlet (red dot number 2) to the basin outlet (red dot number 1). (\textbf{c}) represents the moraine in spring at its snow maximum accumulation while (\textbf{d}) exhibits the snow free moraine in autumn. These 2 photos were taken from one of the highest point of the basin at app. 800 m.a.s.l. {\color{red}The red line on the oblique view images \textbf{c} and \textbf{d} matches the yellow line on the aerial photography \textbf{b}, providing scale with respect to the orthorectified aerial picture.}}\label{2}
\end{figure}

This work was carried out on a small glacial basin located on the West coast of Spitsbergen (high-Arctic), on the North side of the Br\o gger peninsula (79$^\circ$N, 12$^\circ$E, Fig. \ref{2}).

{\color{black}With} a $10.58$~km$^{2}$ basin, Austre Lov\'en (AL) is a small land-terminating valley and polythermal glacier that covers an area of $4.5$~km$^{2}$, with a maximum altitude of no more than $550$~m.a.s.l. AL exhibits a strong negative mass balance with a mean ablation rate of 0.43~m.a$^{-1}$ between 1962 and 1995, which increased to 0.70~m.a$^{-1}$ for the 1995–2009 period, as  reported by \cite{Friedt2011}. As many small glaciers around, AL is surrounded by rugged peaks and slopes that stand out against a flat forefield where surface run-offs are very dynamic (\cite{Marlin2017}). The moraine is today a $2.4$~km$^{2}$ large sedimentary complex which was formed since the Little Ice Age (LIA) period, around 1860 in this region.  
Hence, {\color{black}the moraine} exhibits {\color{black} features representative of} successive retreats of the glacier with a particular shape at the interface with the glacier snout, due to the fast retreat during the last decade (\cite{Kohler2007a}). The combination of glacier melting, temperature variability and increasing precipitation (\cite{Hock2007}) widely favor processes such as sediment transfer (\cite{Bernard2018}), melting and runoffs (\cite{Hagen2005}). Under these dynamics, the proglacial moraine constantly reshapes from one year to another due to the glacial retreat exposing brittle material in a rough topography. With such an heterogeneous morphology coupled with a significant geomorphological and hydrological activity, the proglacial moraine is a key area. Indeed, several snowcover processes, such as melting processes (\cite{BernardEric2013}, \cite{Ewertowski2015}) as well its role of water storage (\cite{Wittmeier2015}) play a key role in the broader source-to-sink dynamics.       

\section{Material and methods}
\subsection{Reference data}
We designed our study on reference data on which our measurements are based and compared. The baseline {\color{red}Digital Elevation Model (DEM)} used was obtained from ArcticDEM 
(\url{https://www.pgc.umn.edu/data/arcticdem/}). This model refers to 2015 images and provide a 2~m resolution DEM, accurate enough 
{\color{red}with respect to the natural (boulder) or size of the topographic features where 
artificial GCPs were located for UAV data referencing and validation}. 

Aerial images used as reference were provided by Norsk Polarinstitutt (available as a {\color{red}Web Map Tile Service (WMTS)} service at 
\url{http://geodata.npolar.no/arcgis/rest/services/Basisdata/NP_Ortofoto_Svalbard_WMTS_25833/MapServer/WMTS/1.0.0/WMTSCapabilities.xml}). 
The image that corresponds to the AL area {\color{black}was acquired in 2010} with a resolution of {\color{black}16.5~cm}, well suited for ground control point localization. Overall definition and optical quality of the images were helpful in order to localize and point erratic boulders and terrain features that were used as control points .  

\begin{figure}[h!tb]
\centering
	\includegraphics[width=\linewidth]{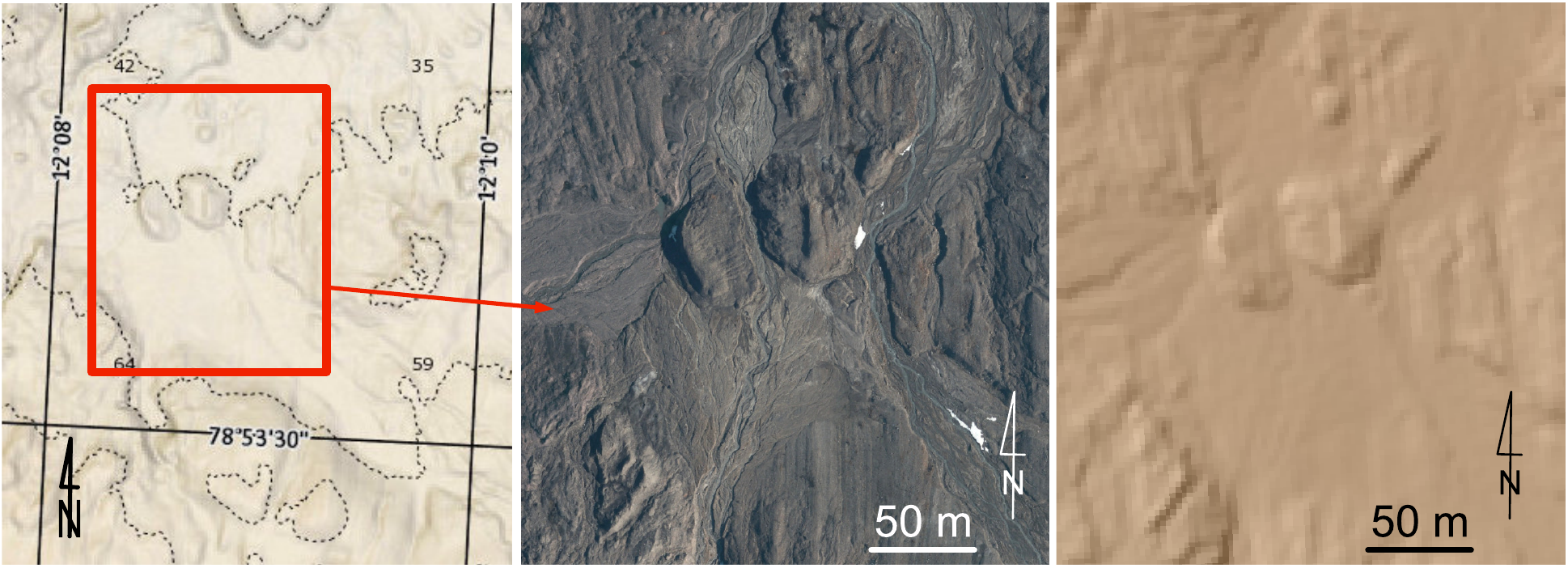}
\caption{Reference data, with (left) the physical topographic map used for basic controls and localization. Center: reference orthophoto fetched on the WMTS server of Norsk Polar Institutt (acquired in 2010), cropped to the region of interest. Right: associated DEM with a 2~m resolution}
\label{3}
\end{figure}

In addition to these data, the Topo Svalbard physical maps were also used as terrain references and implemented into the Geographical Information System software ({\color{red}Quantum Geographic Information System -- QGIS available at \url{https://qgis.org}}) used for this study. Fig. \ref{3} illustrates an overview of these reference data and highlights the focus on the region of interest.  

\subsection{Image acquisition protocol}	
\begin{figure}[h!tb]
\centering
	\includegraphics[width=.8\linewidth]{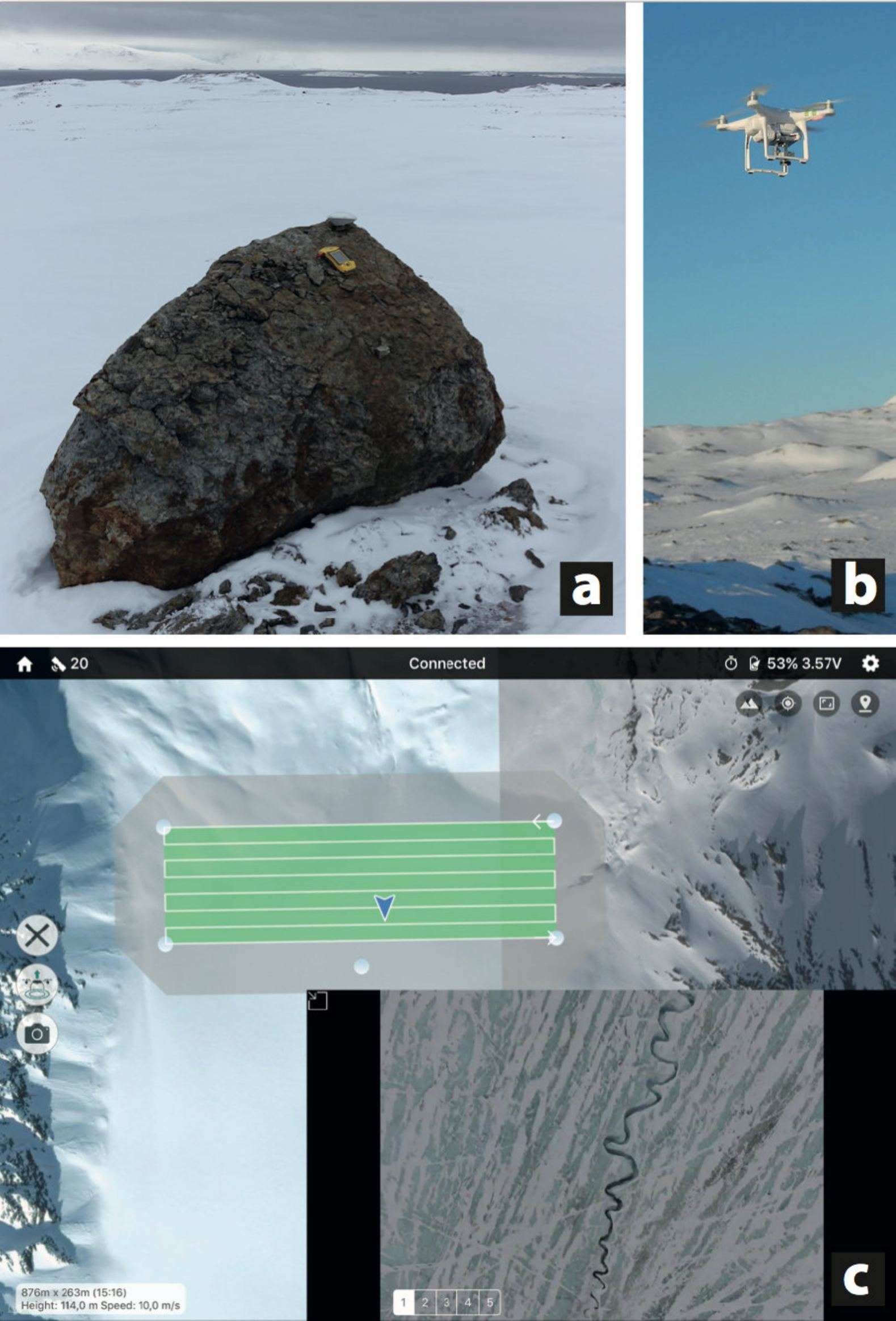}
\caption{Experimental setup: natural Ground Control Points are selected for their good visibility even with the heavy spring snow cover. (a) Their position is measured using a dual-frequency GPS receiver, or identified on a reference orthophoto. (b) A COTS UAV is used for nadir picture acquisition from an elevation of 110~m above take-off altitude. (c) Real time feedback of the camera view and GPS position of the UAV during acquisition improve safety and allow adapting the flight path to features seen in the aerial views. Altizure software is used for setting the raster scan flight path accounting for horizontal speed, image coverage and field of view.}
\label{4}
\end{figure}

Image acquisition was undertaken by using a Commercial Off the Shelf (COTS) DJI Phantom 3 Professional UAV (Fig. \ref{4}) {\color{red}fitted with its on-board camera based on the 1/2.3'' CMOS, 12~Mpixel -- $4000\times 3000$~pixel images -- sensor exhibiting a field of view of 94$^\circ$ as would be obtained with a 20~mm equivalent lens on a 35~mm film.}
The UAV was used with its associated control hardware and software (DJI GO) while keeping original settings, {\color{black}meaning} that the camera was not calibrated. Flight elevation was set at approximately 110~m above ground level, providing a ground resolution of $5\times 5$~cm per pixel considering the optical system lens properties. {\color{red}A total of
2795~pictures were collected in September 2016 during five successive flight sessions covering
the whole moraine and glacier snout, and 1699~pictures were collected in April 2017 over
two days.}

Every camera parameters (ISO, shutter speed and focal aperture) have been manually set, mainly 
depending on the light conditions as well as the ground nature (bare stones, ice, snow) as 
recommended in several works in a similar environment 
(\cite{Westoby2012}, \cite{Lucieer2014}, \cite{Colomina2014}). 

Besides, a dedicated mapping software (Altizure, \url{www.altizure.com}) {\color{black}was used} 
in order to define 
raster-patterned flight plannings and storing such paths for later reproduction. Pre-defined 
flight plans and settings give a systematic approach which improves efficiency and allows
for faithful {\color{black}repetition of the flight path, convenient for} further data processing 
(photo overlap, triggering interval). These plans could also be used again afterwards in order 
to repeat the observations, ensuring a similar protocol of data acquisition. 
{\color{red}The overlap between pictures was set in the Altizure software 
to 80\% in the fast scan flight direction and a sidelap of 60\% in the slow-scan direction.} 
Further details on 
this survey setup and validation were documented by \cite{Buhler2015a}. 

{\color{black}For this} work, data acquisitions were made in autumn (during the most likely snow-free moraine period, beginning of October) and in spring (late April) at the theoretical snow peak accumulation. In autumn 2016, UAV survey was flown during a single campaign in order to get both homogeneous images and a proper light pattern ({\it i.e.} no drop shadow and a sufficient light). In spring 2017, a two-day survey period was necessary because of the impact of cold weather on battery capacity. The total area covered by these surveys is c. 2~km$^{2}$, representing 88~\% of the total moraine surface. The {\color{black} 
investigation is split between a broad analysis of water storage volume computed through the snow cover thickness distribution, in addition to a 
focused investigation on the canyon and icings dynamics on a restricted 0.31~km$^{2}$ region of interest (ROI) {\color{red}as described in section \ref{ROI}}.}

Both periods present technical difficulties. In autumn, {\color{black}grazing sun and} low lights imply careful camera settings ({\color{black}aperture, 
speed} and ISO choice) as well as short measurement intervals. The goal is to prevent cast shadow from inducing excessively variable observation of the same scenery during repeated passes of the UAV over the same region. In spring, the high reflectance, the lack of structures on the smooth snow cover, and low contrast also make photogrammetry challenging. Nevertheless, we observed protruding rocks or snow structures ({\it i.e.} sastrugis) to offer some usable tie points in most moraine areas ({\color{black}Fig. \ref{4}}). 

GCP {\color{red}coordinates} were measured using a dual-frequency GPS receiver (Geo XH device with Zephyr antenna) and post-processed using RINEX data obtained from the EUREF Permanent Network station at Ny-{\AA}lesund (\url{http://www.epncb.oma.be/networkdata/}) 
{\color{red}and their position and elevation cross-checked with the ArcticDEM also used to assess the consistency of the resulting DSM.}

The 2016 and 2017 imagery datasets were georeferenced using 25~GCPs. Twenty of them consisted of erratic boulders easily {\color{black}identified on an aerial picture} since they are at least bigger than $1$~m$^{3}$. The other 5 GCPs consisted of pink plastic gardening saucers targets with a 30~cm diameter placed where no natural GCPs could be identified. They have not been permanently installed and have been deployed each time a few hours prior to the UAV flight. These saucers were also used as reference points. 	 

Two parallel processing flows were run for independent assessment of the error sources:
\begin{itemize}[leftmargin=*,labelsep=5.8mm]
\item in autumn, dedicated GCPs were deployed in the moraine along the flight paths, and their 
position was recorded  prior to the UAV flights. According to post processing, the accuracy obtained 
reached values of 15~cm for 98\% of the markers, in the 3 directions (X, Y, Z). 
\item large boulders were identified on the Norsk Polar Institutt orthophoto used as reference,
{\color{red} and
thanks to the ArcticDEM, the three coordinates of these reference points are identified 
and used as GCPs in addition to positioning on the field using the dual-frequency GPS receiver}.
\end{itemize}

\subsection{Manual snow measurement}	
Unlike the glacier which exhibits a low-roughness surface, the moraine is characterized by a changing and most of all very rugged terrain.  
In such a context and to ensure a reference measurements and then compare with DSM deduced snow cover thicknesses, an avalanche snow probe was used to determine the snow depth. This efficient way to measure quickly snow depth meets the objective of obtaining accurate values on a recurring basis. As reported by several works on snow science (\cite{Grunewald2010}, \cite{Brulandl2001}), it is the most common, the easiest and the most reliable way of snow measurements protocol (especially considering local scale works).
Thus, to assess the quality of our data, {\color{red}50}~probings were carried out by using a 3~m long snow probe with centimetric graduations during the same period as the UAV dataset was collected. A single operator made the campaign to avoid shifts in the way of probing {\color{black}along a transect following the central flowline. It extends over the glacier front to the maximum LIA glacier extent corresponding to hummocky moraine limits}. 
Although probing values cannot be spatially interpolated due to the strong variability given by the uneven ground (which is one of the issues that initiated this work), these points provide a valuable one-off validation dataset which was compared with photogrammetric data. 

\subsection{Data processing}
In this study, we adopted the SfM workflow as implemented in the commercial software package Agisoft PhotoScan Professional version 1.4.5. both for DSM and orthophoto generation. Its efficiency for such purpose ({\it i.e.} geosciences and cold environments) was highlighted in several previous publications (\cite{Cook2013} and \cite{Harwin2015}), and it represented a robust solution to achieve the goals we had set up for this work. 
The detailed description of the SfM procedure using Photoscan is described in \cite{veroheven}: the classical steps for ground surface reconstruction have been followed according to a three-step process, as described and used by (\cite{Dietrich2016} and \cite{Uysal2015}). We used different photo chunks, in order to select regions of interest (ROI) into the moraine. 
More specifically for this work, we have chosen to edit several photos from spring acquisition. At this period, almost all the ground is covered by snow, which gives a texture / colour consistency. This uniformity does not allow the {\color{black}registration} algorithm to work properly unless some features such as rocks or bare stones can be found. To overcome this issue, we {\color{red}edited the photos with the Affinity Photo 
(version 1.8.4) software by using a batch processing, first increasing the contrast (slider at 50\% of the available range) and then the 
sharpness (setting ``70\%'') with the ``high pass sharpening tool''}.


Generated data from Agisoft Photoscan were afterwards processed in QGIS open source software (LT version 2.8). Images were analyzed with the classical raster tools and the Object-Based Image Analysis (OBIA) to obtain surfaces of icings and a representation of the hydrological network. 

In order to analyze the DSMs, we used the SAGA plugin which provides a robust toolbox for geosciences purposes {\color{red}as described at \url{sagatutorials.wordpress.com} whose 
``Terrain Analysis and Processing'' and ``Hydrological Flow Path'' processing flowcharts were 
followed}. In the SAGA toolbox, we first used the ``terrain analysis $\rightarrow$ catchment area'' 
tool in order to determine and apply the same catchment surface of comparison to both DSMs. Then, 
the ``morphometry'' library allowed to correct potential artefacts and close gaps in the DSMs. 
Finally, the last step was completed with the ``raster calculus $\rightarrow$ raster volume'' tool 
to compute the differences between both DSMs ({\color{red}Difference of DSMs}, DoD) and hence to estimate the volume of snow.
These processing steps led to quantify the remaining quantity of snow and the volume of melted snow and residual icing accretion.  
Processing DSM differences over the whole moraine {\color{black}is challenging}. Nevertheless, to assess snowpack accumulation over time, a raster difference layer was created by subtracting the 2016 (October) and 2017 (April) DSMs. 
The entire area recorded was cropped to fit the area of interest. This area includes the outlet at the front of the glacier, following the main 
stream, up to the external moraine. This sequence represents the most rolling and changing topography.

\section{Results and discussion}
\subsection{Morphological evidence of icings spatial dynamics}
\begin{figure}[h!tb]
\centering
	\includegraphics[width=.9\linewidth]{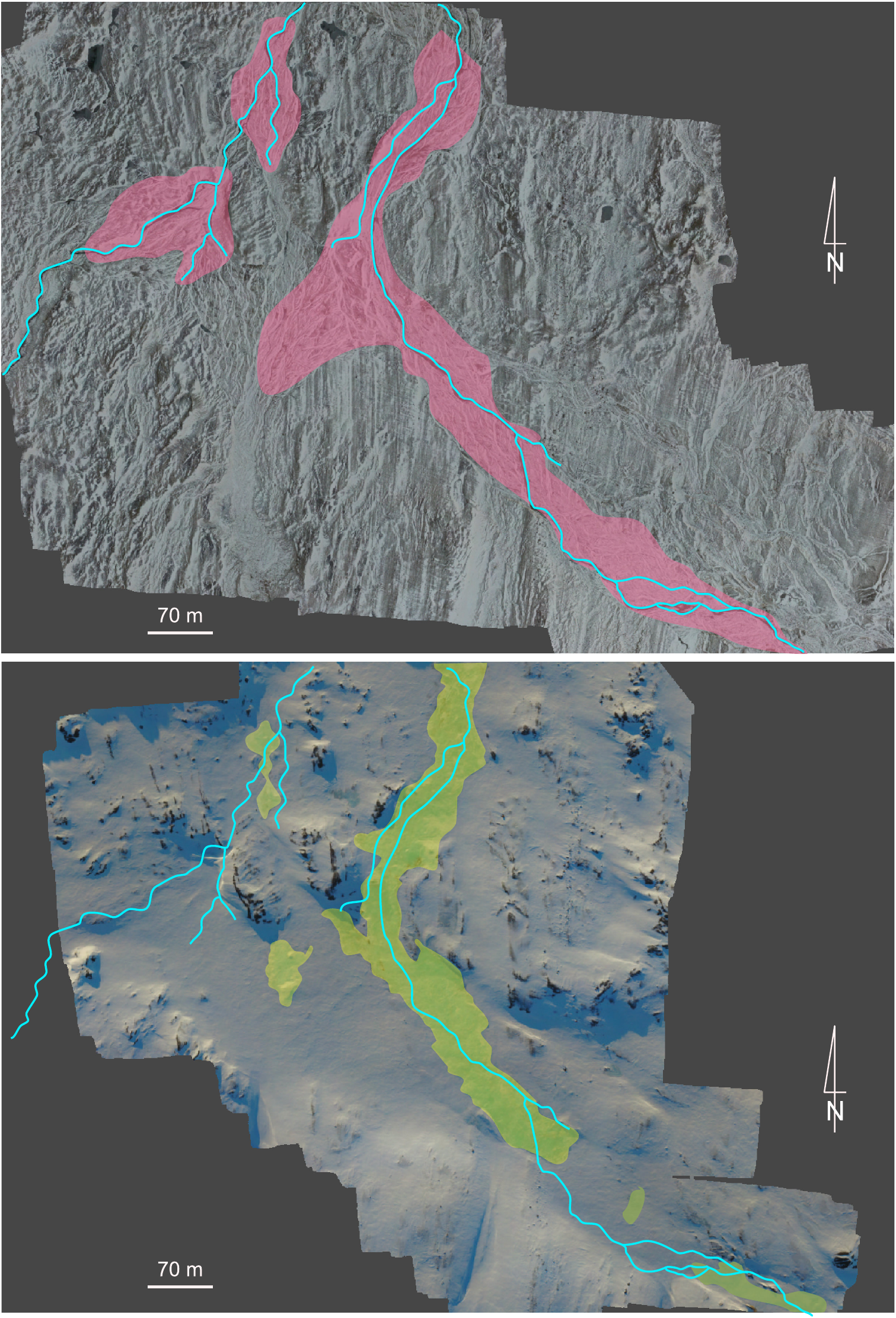}
\caption{Top: localization of the remaining hydrological network ({\color{red}light blue}) coming from the melting season, and mapped on the ortho-image of October. These are the main channels where the major part of run-offs occurs. {\color{red}The pink areas exhibit the maximum extent of the icings at the beginning of
the melting season.} Bottom: the hydrological network ({\color{red}light blue}) is overlapped on the ortho-image acquired in April, {\color{red}with the yellow
areas highlighting the icings present at the time of acquisition}.}
\label{marges}
\end{figure}

The analysis of orthoimages shows significant differences on icings size and distribution between the maximum snow accumulation and the end of the hydrological year 
({\it i.e.} October to September {\color{black}of the next year}) (Fig. \ref{marges}). During the last years, in the moraine, {\color{red}we have observed firn areas getting} smaller or even completely disappearing during the melting season. In this example, the surface of icings varies between a maximal extent of $0.087$~km$^{2}$ to a residual extent of $0.015$~km$^{2}$ at the very end of the hydrological season.  

Localization of remaining icing structures at the beginning and at the end of the season demonstrates the active water {\color{red}upwelling by capilarity} through the snowpack as explained by (\cite{Bukowska-Jania2007}). Especially in the period of maximal snow accumulation, processes are not fixed and a huge amount of liquid water flows into the snowpack depending on its quality ({\it i.e.} hard pack vs fresh snow). 

In autumn, remaining icing areas are mainly located in the rugged part of the proglacial moraine where the impact of radiation is the lowest and where the heavier cold katabatic air preferentially flows. In the case of Austre Lov\'en basin, this means that icings are essentially located on the right bank (East side) of the proglacial moraine. These old canyons concentrate most of the firn accumulation which persists over a hydrological year. This situation contrasts with active periods, in spring. Icing field localizations correspond to the stream bed of the main outlet, but include a large part of its floodplain. Indeed, these are areas where the snowpack is less thick: combined with the action of strong pressure, the liquid water reaches the surface. This results in a wider area, which evolves very quickly from one day to another as mentioned by \cite{BukowskaJania2005}. This is a point that we observed on the field and which is actually impossible to map: UAV flight sessions should be carried out at least every half a day to highlight such dynamic changes. If we compare images acquired in April 2017 with older data (satellite images from 2007-2009), icing fields today are less fragmented, but much wider compared to previous year observations. 

In the active area, {\it i.e.} the main proglacial river, the shapes of icings are more complex and elongated than in the inactive area. Moreover, while the inactive area exhibits residual icings, dynamics along the main rivers are more complex. During the melting season, the part of the icing spreading in the river channel usually melts completely. We observed that, into the proglacial moraine, flat proglacial zones favor the formation of larger icing fields. As already described, the snowpack seems to play a significant role in the development of icing mounds. The water which flows out of a glacier moves in and on the snowpack until {\color{red}upwelling by capilarity}
is stopped with sub-zero air temperatures. In the case of Autre Lov\'enbreen proglacial forefield, the compact structure of canyons as well as snow accumulation block the water which accumulates and flows out under pressure.   

\subsection{Data quality assessment: snow depth calculation}\label{ROI}
We aim at determining the accuracy of remotely measured snow depth {\color{black}with respect to manually probing the snow cover thickness, considered as the reference method}. We applied a Bland and Altman test (\cite{Bland1986}) on reference values extracted from the manual probing transect and on the corresponding values to be tested and given by the DoD. The heterogeneity of the measurements is assessed thanks to the varied terrain crossed by the transect (from a rugged and complex topography to a flat smooth ground). 
{\color{red}The consistency of the results obtained by the two measurement methods -- avalanche
probe and DoD -- is assessed with the estimate of the mean bias and standard deviation error
between the two datasets.}
As described by \cite{Bland1986}, we calculated a confidence interval of 95\% which gives the Limits of Agreement (LoA) also derived as mean value $\mu\pm 1.96\sigma$ with $\sigma$ the standard deviation. 

\begin{figure}[h!tb]
\centering
	\includegraphics[width=\linewidth]{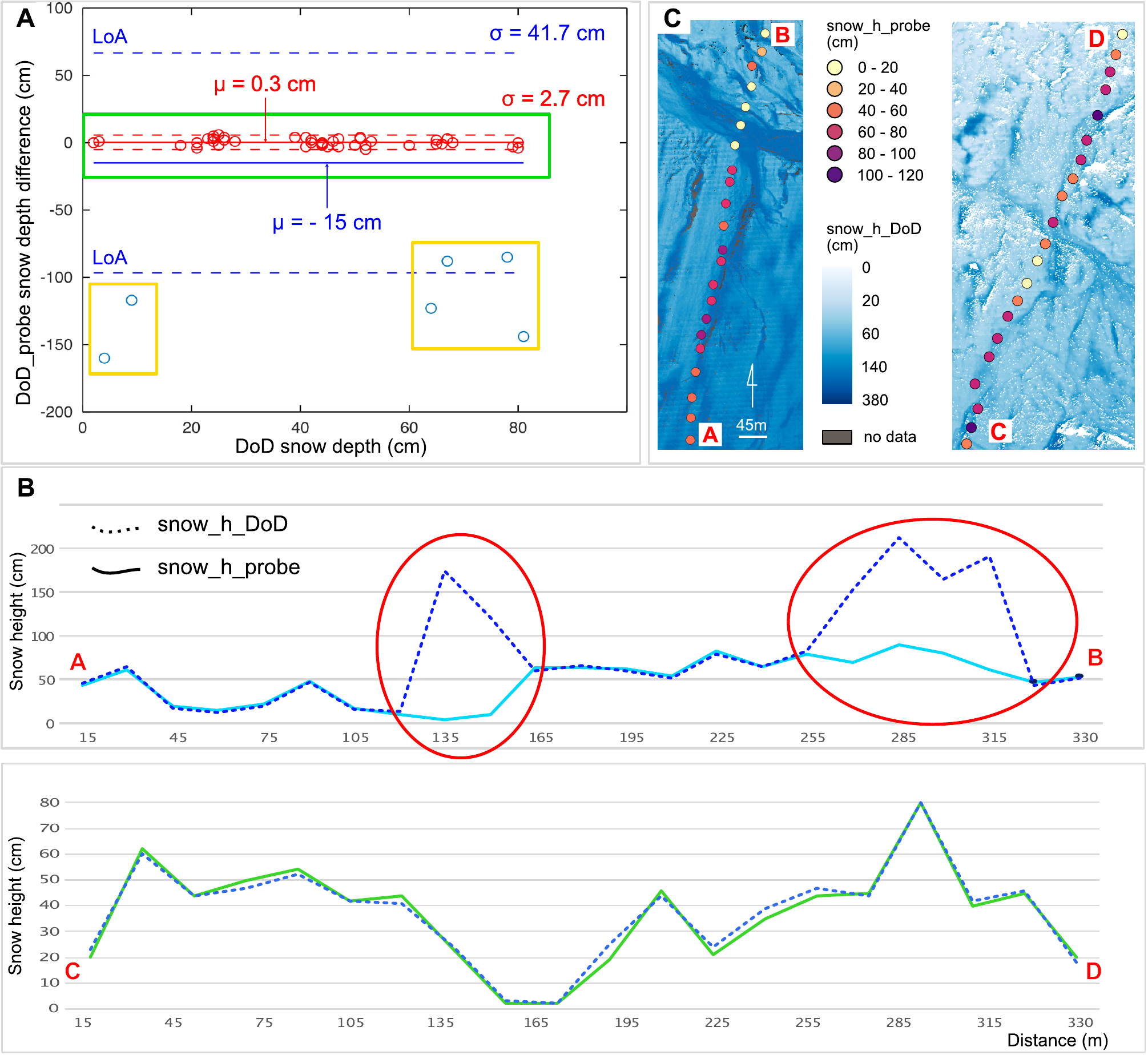}
\caption{
Two cross sections (C) were selected {\color{black}for comparing DoD and manual avalanche probe measurement snow depths} according to their geomorphological features as well as its dynamics: one section crosses icings areas (A--B) while the other is known to only include rock-covered areas (C--D). (B) displays the error between both
measurement techniques as a function of snow depth, emphasizing the low error bar (less than 3~cm) and bias (less than 1~cm) between measurements using the avalanche probe v.s DoD when eliminating measurement over icings (red), while the bias (mean value $\mu$ as solid line) increases to -15~cm and the Limit of Agreement (LoA, dashed lines) equal to 1.96~times the
standard deviation $\sigma$ to nearly 42~cm (blue). (A) and (B) show that the inconsistencies between avalanche probe measurements and DoD only correspond to the icing areas {\color{black}filled with varying ice thickness depending on} the season. (A) highlights the test of Bland and Altman which validates the consistency of the method used ({\it i.e.} difference of DSMs) with respect to a reference method (DoD)}
\label{l}
\end{figure}  

Results of this test are reported on Fig. \ref{l} and highlight an excellent agreement with an average of less
than 1~cm difference between the reference method (manual avalanche probing) and the tested method (photogrammetry and DoD) whenever the snow covers rocky areas and as long as icing fields are not crossed.
All outliers (squared in yellow on {\color{red}Fig. \ref{l} (A)}) correspond to areas where icings dynamics occur (red circles in Fig. \ref{l} (C)) and from which the differences of measurements are associated with the presence of ice. However, the CD section, which is located on an icing free area, shows no shift and thus a significant convergence between both methods. Not removing these erroneous measurements point sets would raise the bias to $-15$~cm and
the LoA to 42~cm, emphasizing the need to mask out icings when processing data collected over the moraine
to establish the snow cover water equivalent volume (section \ref{weq}).

In fact, the avalanche  probe is unable to drill through such compact ice and yet the icings having melt in autumn will add to the DoD thickness measurement. Since icings are localized 
in the moraine to the riverbeds, their contribution to the total snow volume calculation will be negligible. The morphology of these steep-edged riverbeds makes
them prone to be filled with snow, with shapes forcing vertical capillaries upwellings. It is most likely on such morphological shapes that icing processes usually occur. {\color{black}The compact and hard} snow/ice mixture does not allow the probe to reach the ground, explaining the difference of measured values
as further in section \ref{44}.


\begin{figure}[h!tb]
\includegraphics[width=0.49\linewidth]{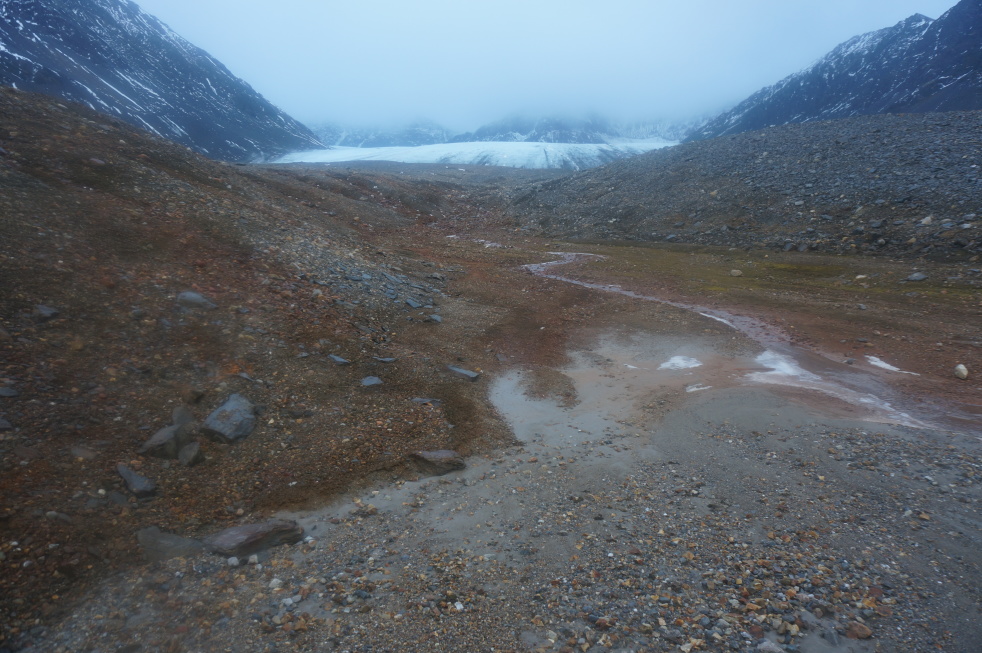}
\includegraphics[width=0.49\linewidth]{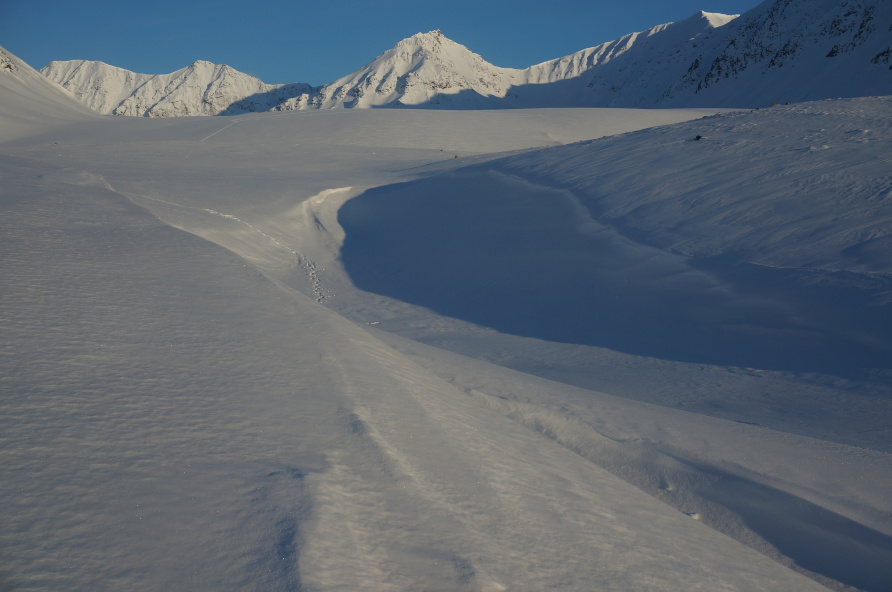}

\includegraphics[width=0.49\linewidth]{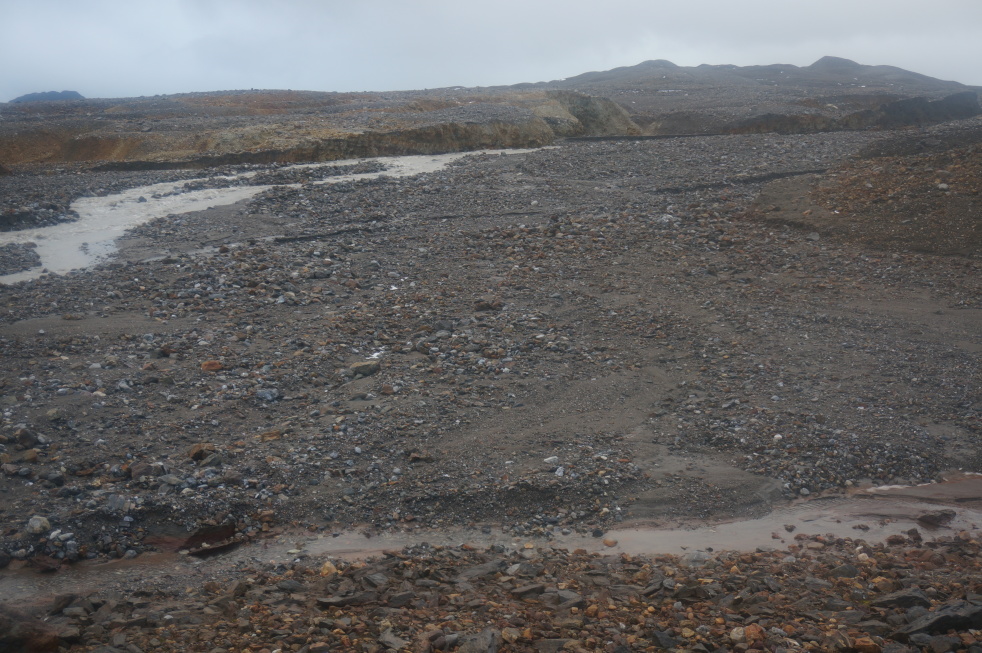}
\includegraphics[width=0.49\linewidth]{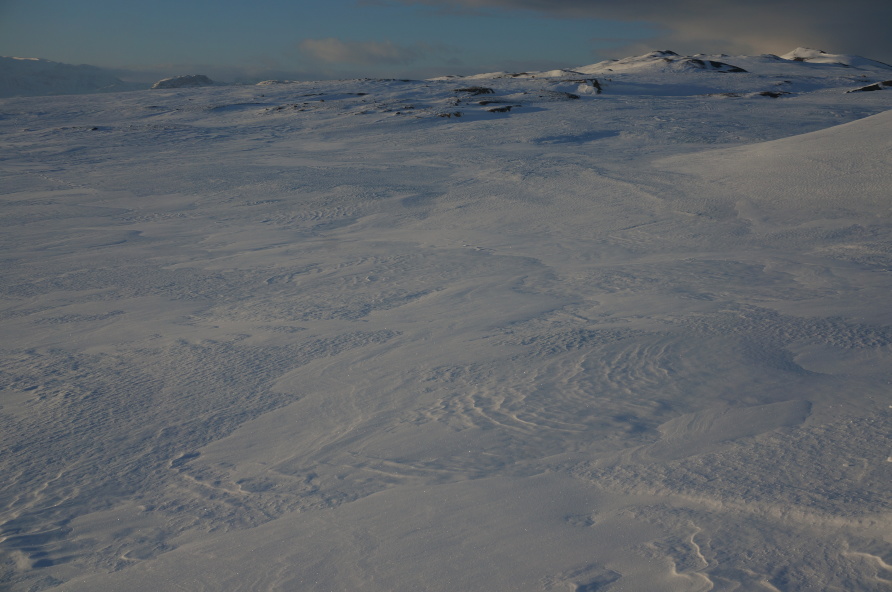}
\caption{Pictures taken from the same spot in April 2017 (right) and in September 2016 (left), towards the glacier (top) or
the fjord (bottom), illustrating the smoothing effect of snow
accumulation and hinting at the consistency of the observed ice and snow thicknesses quantitatively deduced from difference
of DSMs.}
\label{mor2}
\end{figure}

Difference of DSMs applied to snowpack {\color{black}thickness measurement emphasizes} the importance of ground topography. Fig. \ref{mor2} highlights the smoothing effect (images below) when landforms exhibit very small {\color{black}topography}. 
On the contrary, when the landforms are quite sharp (Fig. \ref{mor2}, top images), even a strong wind effect 
is unable to smooth the surface. 
Indeed, a rugged topography promotes cornices formation. This is a much easier configuration for snow depth estimation since:
\begin{itemize}[leftmargin=*,labelsep=5.8mm]
\item the snowpack is {\color{red}decimeter to meter} deep and so easier to estimate by using photogrammetry ;
\item during data processing step, cornices create shadows and structures that are identifiable by processing algorithm.
\end{itemize}

\subsection{Water equivalent calculation}\label{weq}

One of the main topics of studying a small glacial basin is to better understand melting processes and their interaction with 
climate. As was demonstrated \cite{Bukowska-Jania2007}, icing fields constitute a very important element of the cryosphere in the 
High Arctic as a witness of thermal transformation of the glacier, and thus indirectly on the climate change impact. 

Measuring snow water equivalent (SWE) requires substantially more effort than only sampling snow depth (HS). SWE and HS are known to be strongly correlated (\cite{Deems2013}). This correlation could potentially be used to estimate SWE from HS even with few sampling points.
Thus, studies have suggested enhancing sampling efficiency by substituting a significant part of the time-consuming SWE measurements by simple HS measurements (\cite{Debeer2009}). In our case, we carried out some snow sample measurements while flying the UAV, ensuring data acquisition at the same time. Snow samples were collected in snow pits at depths ranging from 20 to 100~cm by using 125~ml plastic bottles. 
Snow cover thickness was measured by using an avalanche probe following the same protocol as reported before. Despite varying snow conditions in various areas 
of the moraine, depending on the surrounding topography yielding more variable snow conditions than on the smooth glacier surface, the snow density was found to be homogeneous and constant at 0.43$\pm$0.03 {\color{red}relative to water (1~g/cm$^3$)}. This value is equal to those typically observed around the peninsula (\cite{Obleitner2004} and \cite{Winther2003}).

\begin{figure}[h!tb]
\centering
	\includegraphics[width=\linewidth]{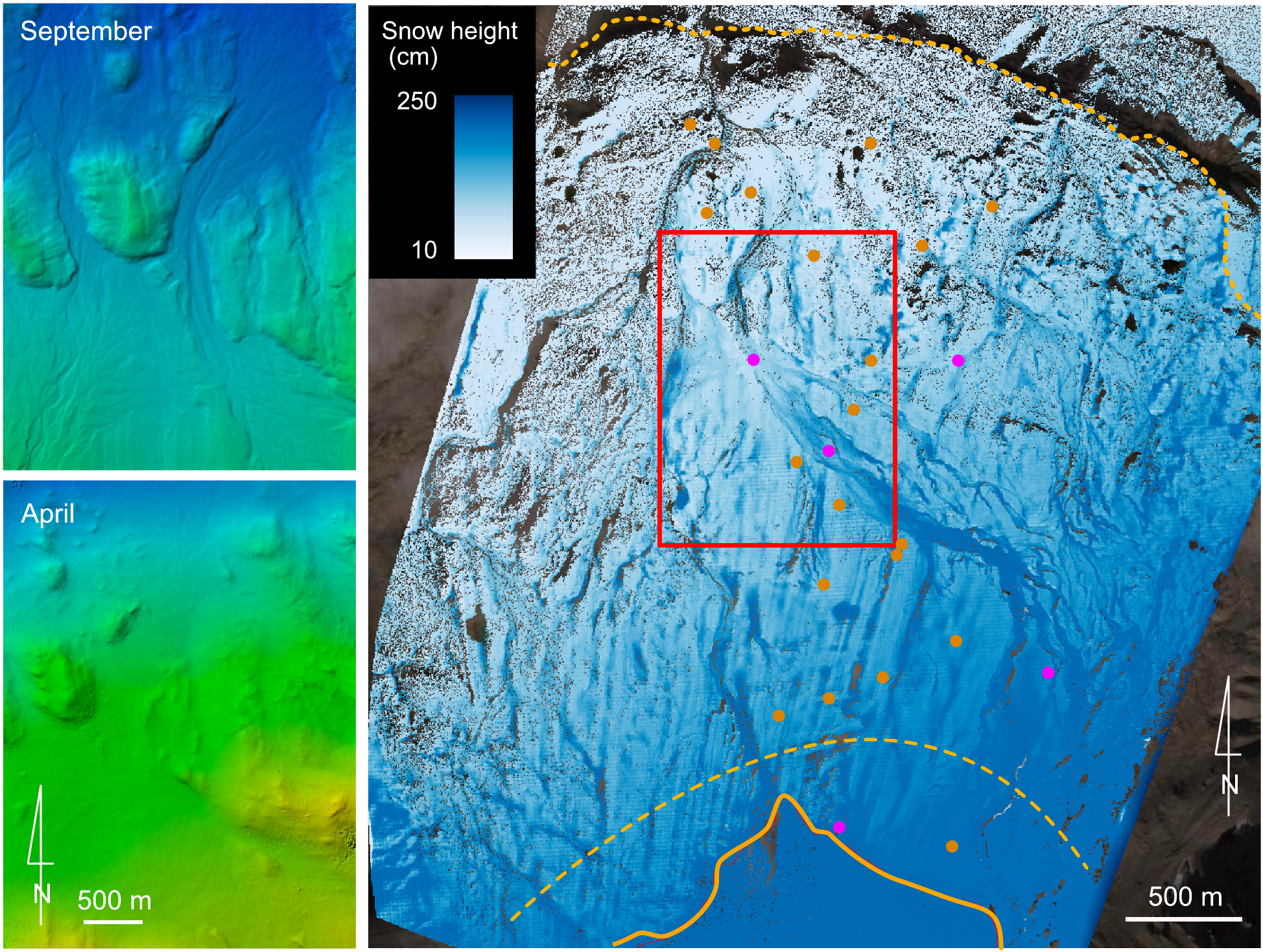}
\caption{DSMs were generated for both campaigns in October 2016 and April 2017. The {\color{black}spatial} resolution was set at approximately 50~cm for both DSMs. 
The resulting difference of DSM is depicted on the right, with a threshold level at max. 250~cm. The yellow dotted line is the maximum extent of the moraine during LIA, 
the orange solid line is the current glacier front limit and the dashed line corresponds to the limit of the new deglaciated area ({\it i.e.} during the last 10 years). {\color{red}The dots indicate the location
of the GCPs, with red dots for natural GCPs consisting of large ($>1$~m) boulders, and pink dots
for artificial GCPs located in flat areas. The red rectangle on the right image provides a geographical setting of the 
Region of Interest highlighted in both left images.}}
\label{m}
\end{figure}  

Snow depths deduced from DoD as shown in Fig. \ref{m} are used for water equivalent estimate over the whole moraine
area. {\color{red}The mean snow thickness in the 2.2~km$^2$ area of the internal moraine deduced from DoD is 333~mm, 
which multiplied by the snow density of 0.43 leads to 143~mm.SWE. This measurement excludes the hummocky moraine with its 
morphology characterized by a convex shape leading to rather snowy conditions. Our measurements on the 4.5~km$^2$ 
glacier indicates a snow contribution of 491~mm.SWE. Hence, the snowpack contribution of the moraine accounts
for $143/491=29$~\% of the glacier contribution. Normalized to the whole $10.56$~km$^2$ basin, 
the moraine SWE contribution is $143\times 2.2/10.56=30$~mm.SWE which compares to the glacier contribution
of $491\times 4.5/10.58=209$~mm.SWE or a relative contribution of 14\% normalized to the whole glacier basin. 
This statement should be balanced since only the snowpack is included in this estimate while
groundwater and run-offs due to liquid precipitation are not taken into account in this calculation.}

{\color{red}The error budget is as follows: since the SWE is deduced from a product of the density $\rho$ with the thickness $h$
of the snow cover, the uncertainty on this quantity is $$\frac{dSWE}{SWE}=\frac{d\rho}{\rho}+\frac{dh}{h}$$ where $dx$ indicates
the uncertainty on quantity $x$. Here $h=333$~mm mean snow cover height so that the relative uncertainty $dh/h=27/333=0.08$
according to the analysis of Fig. \ref{l}, while the density uncertainty contributes to $d\rho/\rho=0.03/0.43=0.07$. Thus, both
quantities contribute equally to a total uncertainty of 15\% on the SWE.}

These values emphasize the importance of the snowpack stored in the moraine in the hydrological equation of the watershed. This 
quantity of snow partly explains the increase of water runoffs at the melting season. The potential release of 
a massive amount of water increases sediment transfer as observed in \cite{Bernard2018}. {\color{red}This analysis solves one 
of the missing variables in the hydrological equation which includes the glacier area, here the moraine area and the still missing 
slope contribution to the global hydrological budget}.

\subsection{Lessons learnt and outlook}\label{44}
The estimation of snowpack characteristics is challenging by using SfM photogrammetry. However, the measurements performed on the 
proglacial moraine assessed that there are strong snow drift effects. Regardless of snow accumulation, it appears that 
morainic mounds evolve very little, contrary to canyons, that are constantly re-shaping and subject to strong melting 
processes that consequently dig under sediment transport action. Thus, the structure of the topography promotes massive 
snow accumulation as well as the orientation, orthogonal to the dominant winds. 
A lesson learnt while studying snowpack in the moraine, is that the comparison with the glacier snowcover is possible with a low residual uncertainty but requires two workflows. 
Previous works (\cite{Gindraux2017}, \cite{Buhler2015a}) showed that on the glacier a simple interpolation can be applied to estimate both SWE and height. In the case 
of the proglacial moraine, the difference of DSM calculation is recommended since an interpolation is not consistent with its rugged topography.
As often observed, the moraine constitutes a 
key area but still hard to monitor. Based on this paper and previous works, the coupling of LiDAR measurements as references, 
and several photogrammetric flight session appears as the most efficient method. 

{\color{red}UAV airborne data acquisition appears as an efficient vector
when addressing an investigation area of a few km$^2$ with data collection lasting
less than half a day, meeting the assumption of static measurement conditions. 
Photogrammetric SfM processing has then been used for generating DSM whose difference
led to various geomorphological and snow cover evolution characterizations.}
Considering wider investigation areas, the uncertainty is more important and recent works carried out on the same area using 
{\color{red}RAdiofrequency Detection And Ranging (RADAR)} leads to convincing results. However, results given by RADAR are strongly correlated with the {\color{red}types} of the snowpack: the presence of ice layers (due to rain on snow event far instance) decreases the accuracy of measurement due to the physical properties of RADAR signals and the complex interaction
of electromagnetic waves with the snow pack.
Thus, both methods seem to be complementary: the wide scale approach gives an overview, and data obtained through photogrammetry allow for surveying regions of interest.
In addition, the use of combined UAV campaign and photogrammetry processing is relevant in the context of phenomena {\color{red}occurring with
hourly to daily span with long lasting consequences such as heavy snow fall, rainfall inducing canyon carving and rain on snow events}.  

\begin{figure}[h!tb]
\centering
\includegraphics[width=.9\linewidth]{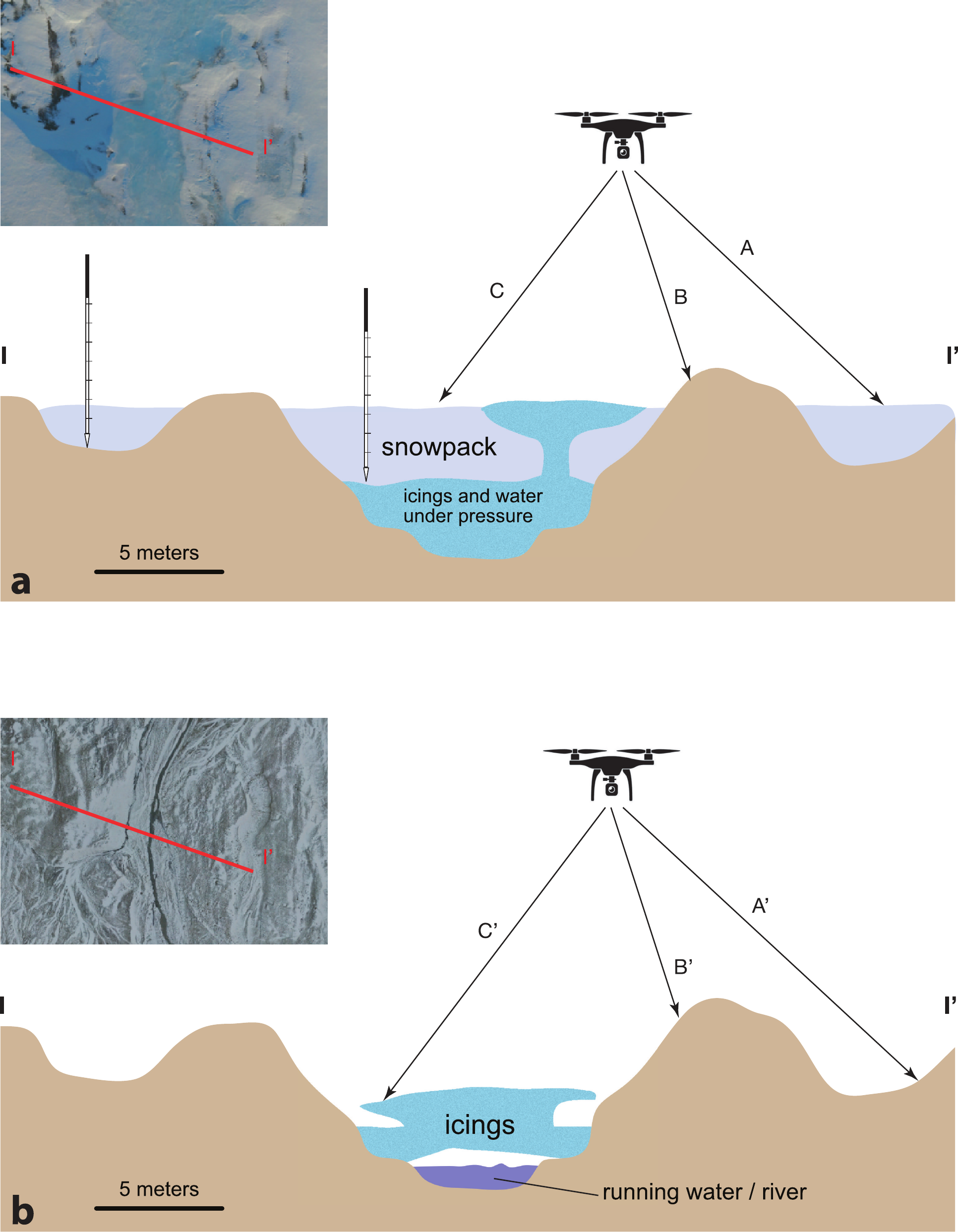}
\caption{{\color{black}Analysis of the} different conditions met by subtracting DSMs {\color{red}collected when the snow
cover is maximum (top) and has melted (bottom). A-A' allows for snow cover thickness measurement, while B-B' provides an
estimate of the accuracy of the measurement by comparing snow-free areas where bare morainic rock is visible. Due to varying
icings disposition, size and volume between the beginning and the end of the season, C-C' computed as the subtraction} of one DSM 
to another does not yield the snow cover thickness. {\color{red}The scale on each cross-section schematic matches the length of 
the red line in the inset pictures. The avalanche probe schematic on the left of the top picture aims at illustrating how
the snow cover thickness is measured over rock-covered area but how the measurement might be biased over icings with
dense ice layers between the snow cover and the bedrock, as is the case of SfM in the C-C' condition.}}
\label{9}
\end{figure}

{\color{black}Discrepancies were highlighted between manual measurements of snow depths and DoD analysis, interpreted as illustrated in
Fig. \ref{9} with the different quantities measured by both methods. Manual avalanche probe snow thickness is limited to the soft snow
layer and does not include the compact underlying ice of the icings. On the other hand, DoD integrates both quantities since it refers
to the ice-free moraine rocky surface observed in autumn. Beyond these icing areas, the snow cover thickness comparisons have been
observed to match with sub-decimeter accuracy, with DoD providing a high spatial resolution that cannot be matched with manual
avalanche probe measurement which cannot be interpolated in the rough moraine area.}

About icings dynamics, the seasonal approach described in this work will need to be extended to several years to better understand 
how its mechanisms are influenced by climate. Nevertheless, inter-seasonal observation gave quite a few lessons. First, 
the presence or absence of icings indicates changes in the functioning of the proglacial moraine internal drainage system. 
Obviously, it appears that icings are not located in the same area in spring and in autumn. But the important point is that in 
autumn, there is no significant dynamics recorded contrary to spring where changes can be observed from an hour to another. 
The spatio-temporal scale at which processes are carried out is too fast to be measurable, even by using UAV surveys. 
It was quite easy to observe the fast formation of massive icing mounds which raises questions about the icings dynamics. 
In autumn the absence of any movement could be attributed to the fact that these icings are no longer in activity nor supplied 
by water outflows. According to \cite{Dietz2013}, this means, in the case of Austre Lov\'en proglacial moraine, that 
almost all icings are associated with rivers, glacial water outflows 
and groundwater outflows.  
This conclusion is supported by spring observations, which clearly indicate the strong relationship between outflows and icings. 

\section{Conclusions}

Two years of snow cover in an Arctic proglacial moraine area were investigated using difference of Digital Elevation Models,
referring to the snow-free dataset acquired in autumn. While spatial correlation is observed with respect to avalanche probe measurements
in areas where snow accumulation over bare moraine rock is significant, the poor general correlation between in-situ measurement and
remote-sensing techniques is attributed to the ice accumulation underlying the snowpack. This result is most striking in icings
areas. Fine digital elevation model registration for snow cover thickness estimate requires ground-based control points. When lacking
artificial reference points, natural ground control points were here used to register past and present acquisitions, referring to 
large boulders clearly visible even at maximum snow cover and known not to have moved in the last 7 years with respect to the reference
orthophoto. Despite poor contrast under homogeneous snow cover conditions, Structure from Motion photogrammetric analysis appears suitable
for mapping snow cover distribution even in the low-lying sun, cast shadow met in Arctic environments.

Mapping a 2.4~km$^2$ area proglacial moraine snow cover characteristics appears 
beyond the reach of a rotating wing quad-copter UAV: estimate SWE for the whole moraine is not
possible with the current dataset acquired over multiple flight sessions due
to the limited (20~minutes at most) autonomy. We conclude that a rotating wing UAV quadcopter is 
not suitable for such a large area. A fixed wing UAV seems to be a better suited solution as 
demonstrated by \cite{Gindraux2017} in which a 5~km$^2$ tongue of a glacier was mapped, an area similar
to the one under investigation here, through flights spanning about 0.35~km$^2$ each, an area about
1.5 to two times larger than those covered during our rotating wing UAV flights. Despite similar flight
elevation and adjacent image coverage, their flight duration at 2500~m.a.s.l is about twice the one we met in Arctic conditions
of close or sub-zero temperatures at sea level (15~minute flight durations for the DJI Phantom 3). 
In addition, combining SfM methods 
with satellite RADAR images analysis will open new opportunities for snowpack study in harsh condition 
as well as in rough topographic environment, thanks to the high resolution DSM generated
by the former technique needed for interferometric analysis of the latter. Despite the poorer
RADAR spatial resolution (5~m for Sentinel 1) and high operating frequency (C-SAR at 5.4~GHz or
a 5.5~cm wavelength) inducing more complex interaction of the electromagnetic wave with the snow cover 
than an optical signal, such a technique \cite{Nagler-2000,Luojus-2007} appears worth investigating
in complement with DSM generated by UAV.

\vspace{6pt} 


\authorcontributions{All authors participated equally to field trips, experiment planning, data analysis and manuscript redaction.}

\funding{This research was funded by R\'egion Franche Comt\'e grant and  the logistic of French Polar Institute (IPEV). \url{https://search.crossref.org/funding}, any errors may affect your future funding.}

\conflictsofinterest{The authors declare no conflict of interest. The funders had no role in the design of the study; in the collection, analyses, or interpretation of data; in the writing of the manuscript, or in the decision to publish the results'.} 





\end{paracol}
\reftitle{References}
\bibliography{RS_template_redac}
\end{document}